\documentclass{article}[12pt]
\usepackage{graphicx}
\usepackage{hyperref}
\usepackage{amssymb}
\usepackage{multirow}
\usepackage{caption}
\usepackage{subcaption}
\usepackage[left=1.5cm,right=1.5cm,top=1.5cm,bottom=1.5cm]{geometry}
\usepackage{authblk}
\usepackage{amsmath}
\begin{document}
\title{Thermodynamic invariance of the energy-momentum tensor under matter-Lagrangian choices and its astrophysical implications in $f(R,T)$ gravity}
\author[1]{Debadri Bhattacharjee\thanks{debadriwork@gmail.com}}
\author[2]{Pradip Kumar Chattopadhyay\thanks{pkc$_{-}76$@rediffmail.com}}
\affil[1,2]{IUCAA Centre for Astronomy Research and Development (ICARD), Department of Physics, Cooch Behar Panchanan Barma University, Vivekananda Street, District: Cooch Behar, \\ Pin: 736101, West Bengal, India}
\maketitle
\begin{abstract}{The correct choice for the matter Lagrangian $(\mathcal{L_{M}})$ in the framework of $f(R,T)$ theory of gravity, has been a fundamental yet often overlooked ambiguity. It has been a long-standing issue, whether to choose $\mathcal{L_{M}}=p$ or $-\rho$ as the proper definition of matter sector. In this work, we show that both choices lead to the same energy-momentum tensor, from thermodynamic point of view. However, for these two choices, the structure of the TOV equations are different. We construct and solve TOV equations using MIT bag model equation of state for $\mathcal{L_{M}}=p$ and $-\rho$, and study the impact of the choices for matter Lagrangian on the maximum mass limit as well as M-R plot of compact stars. It is interesting to note that allowed range of gravity-matter coupling coefficient $(\alpha_{c})$ is also different for $\mathcal{L_{M}}=p$ and $\mathcal{L_{M}}=-\rho$, i.e., $\alpha_{c}$ can not be taken arbitrarily. Notably, through $\mathcal{L_{M}}=p$, we achieve a maximum mass of $2.78~M_{\odot}$, whereas for $\mathcal{L_{M}}=-\rho$, we obtain a maximum mass of $2.41~M_{\odot}$. So, despite the same energy-momentum tensor for different choices of $\mathcal{L_{M}}$, the upper limit of maximum mass is significantly modified.}
\end{abstract}
\section{Introduction}\label{sec1} 
Recent advancements in astrophysical and cosmological observations have provided a deeper and a more comprehensive understanding of the Type Ia supernovae studies \cite{Riess}, determinations of mass density and the energy density associated with the cosmological constant \cite{Perlmutter}, high-resolution measurements of the cosmic microwave background (CMB) \cite{Bernardis}, and observations of CMB anisotropies \cite{Hanany}, alongside theoretical investigations into the cosmological constant and dark energy \cite{Peebles,Padmanabhan} that collectively support the evidence for the current accelerated expansion of the universe and the presence of dark matter. Moreover, the detection of gravitational waves (GWs) by LIGO-Virgo-KAGRA (LVK) network, has opened up new frontier in studying the complex binary mergers \cite{Abbott,Abbott1,Abbott2}. These new findings strongly motivate the necessity of modified gravity theories as potential alternatives or extensions to General Relativity (GR). In this context, numerous theoretical modifications have been proposed as extensions of GR, such as $f(R)$, $f(T)$, $f(G)$, $f(R,T)$, and $f(G,T)$ gravity theories. Among these, the most straightforward and widely studied framework is the $f(R)$ theory of gravity \cite{Nojiri}, in which the standard Einstein-Hilbert action is generalised by replacing the Ricci scalar $R$ with an arbitrary function $f(R)$. The $f(R)$ formalism has significantly advanced our understanding of galactic dynamics and the late-time cosmic acceleration \cite{Capozziello,Borowiec}. Subsequently, Harko et al. \cite{Harko} extended this model by incorporating the trace of the energy-momentum tensor $T$ as an additional argument of the function, thereby introducing the $f(R,T)$ theory of gravity. In their seminal work, the authors demonstrated that the dependence on $T$ could be interpreted as an indication of quantum effects or the presence of exotic imperfect fluids \cite{Harko1}. This framework leads to a non-vanishing covariant divergence of the energy-momentum tensor, introducing an additional force term arising from the matter-geometry coupling, which consequently causes test particles to deviate from geodesic motion. A noteworthy feature of the $f(R,T)$ formalism is its potential to provide a classical representation of certain quantum gravitational effects. From a cosmological perspective, $f(R,T)$ gravity has been employed in various contexts, including holographic dark energy, solar system tests, anisotropic cosmologies, and non-equilibrium thermodynamics. Nevertheless, probing such modified theories through relativistic stellar models remains crucial for assessing their physical viability, as compact objects in strong-field regimes can serve as excellent laboratories to discriminate among different gravitational frameworks. Motivated by these considerations, the present work focuses on the $f(R,T)$ theory of gravity as a representative member of the broader class of GR extensions.

Extensive research has been devoted to exploring various aspects of modified theories of gravity. In recent years, there has been a notable rise in the application of $f(R,T)$ gravity to relativistic stellar modeling, particularly for the functional form $f(R,T)=R+2\alpha_{c}T$ \cite{Harko}, where $\alpha_{c}$ denotes the gravity-matter coupling constant. Within this framework, Moraes et al. \cite{Moraes4} were among the first to obtain an exact solution of the Tolman-Oppenheimer-Volkoff (TOV) equations \cite{Tolman,Oppenheimer} in $f(R,T)$ gravity and to analyse the corresponding hydrostatic equilibrium conditions. Building on their results, Das et al. \cite{Das} investigated compact star configurations within $f(R,T)$ gravity by employing Lie algebra and conformal Killing vector techniques. Assuming linear dependencies of $f(R,T)$ on both $R$ and $T$, and utilising the embedding class-I approach, Errehymy et al. \cite{Errehymy} developed anisotropic compact star models. Sarkar et al. \cite{Sarkar} examined spherically symmetric, anisotropic stellar configurations in $f(R,T)$ gravity by selecting a suitable form for the metric potential $g_{rr}$. Sharif and Yousaf \cite{Sharif1} explored the dynamical aspects influencing the stability of isotropic, spherically symmetric stellar systems in this theory. In a related study, Yadav et al. \cite{Yadav} proposed a singularity-free solution representing non-exotic compact stars, while Rej and Bhar \cite{Rej} analyzed isotropic stellar configurations using the Durgapal IV metric potential within the same theoretical framework. Kumar et al. \cite{Kumar1}, employing the Buchdahl ansatz, presented an isotropic stellar model in $f(R,T)$gravity. Moreover, Vacaru \cite{Vacaru} extended $f(R,T)$ gravity into the realm of Finsler geometry, while Chowdhury et al. \cite{Chowdhury1} developed a Finslerian extension for anisotropic strange stars and Yashwanth et al. \cite{Yashwanth} investigated Finslerian wormhole solutions within the same framework. Additional studies have examined relativistic compact configurations under $f(R,T)$ gravity using various geometrical and physical settings, including the isotropic Buchdahl sphere \cite{Bhar}, realistic equations of state and gravitational-wave events \cite{Lobato}, anisotropic fluid spheres with the Durgapal-Fuloria metric potential \cite{Maurya}, dynamical behavior under the Tolman metric \cite{Hansraj}, fulfillment of the Karmarkar condition \cite{Prasad}, anisotropic and decoupled Tolman VII models \cite{Azmat}, and embedding class-I approaches \cite{Hansraj1}. Furthermore, several investigations have focused on gravitational waves within the $f(R,T)$ framework \cite{Alves}, higher-order extensions of gravity \cite{Holscher}, different forms of $f(R,T)$ models \cite{Chowdhury2}, and polarisation effects of gravitational waves in $f(R,T^{\phi})$ theory \cite{Khlopov}. 

Interestingly, none of the these works contain the proper justification and calculation for the choice of matter Lagrangian. Majority of the works, prefer $\mathcal{L_{M}}=p$ without the proper justification. Hence, the ambiguity in choosing the proper matter Lagrangian has existed for long. Following the seminal work of Harko \cite{Harko}, researchers have adopted the the fact that both $\mathcal{L_{M}}=p$, and $\mathcal{L_{M}}=-\rho$ produce the same form of energy-momentum tensor. The puzzle in choosing the form of Lagrangian originated from the work of J.D. Brown \cite{Brown}, where the action function describing the perfect fluid distribution is presented. In that paper, the description pertains to the cases where the energy density is a function of particle number density $(n)$ and specific entropy $(s)$, as well as isentropic fluids. Further, in the context of perfect fluids in modified gravity, Faraoni \cite{Faraoni} demonstrated some robust theoretical framework, ultimately concluding that the appropriate form of the matter Lagrangian remains elusive. Moreover, in describing the matter Lagrangian with particular sign convention, Mendoza and Silva \cite{Mendoza} showed that for an ideal fluid, the choice of $\mathcal{L_{M}}=\pm\rho$ should be favoured over $\mathcal{L_{M}}=\pm p$. Recently, Haghani et al. \cite{Haghani} has provided the necessary thermodynamic, and geometric approach to study the energy-momentum tensor for both choices of $\mathcal{L_{M}}$. 

Motivated by the recent shed of light, in the present paper, we aim to provide the in-depth calculation that will substantiate the fact that both the choices of $\mathcal{L_{M}}$ yield the same expression for an isotropic energy-momentum tensor. Further, the choice of $\mathcal{L_{M}}$ has a significant effect on the formation of field equations \cite{Harko}, which in turn impacts the formulation of TOV equations. Hence, our goal is to form the TOV equations for different choices of $\mathcal{L_{M}}$, and obtain the maximum mass-radius plots for both the choices. This comparative study may act as a favouring agent for the suitable choice of $\mathcal{L_{M}}$. 

The paper is organised in the following manner: we introduce the basics of $f(R,T)$ gravity in section~\ref{sec2}, and obtain the general form of the modified field equations. Now, we obtain the expressions for the energy-momentum tensors for both $\mathcal{L_{M}}=p$, as well as $\mathcal{L_{M}}=-\rho$ in section~\ref{sec3}. Section~\ref{sec4} deals with the formation of TOV equations in both the approaches, and the concerned M-R plots are illustrated in section~\ref{sec5}. Finally, we highlight the main findings and summarise the work in section~\ref{sec6}.  
\section{Mathematical framework of $f(R,T)$ gravity}\label{sec2}
In the framework of $f(R,T)$ gravity, the action takes the form \cite{Harko}:
\begin{equation}
	\mathcal{S}=\frac{1}{16\pi}\int\sqrt{-g}f(R,T)d^{4}x+\int\sqrt{-g}\mathcal{L_{M}}d^{4}x, \label{eq1}
\end{equation}
where, $f(R,T)$ is a function of Ricci scalar $(R)$, and the trace of the energy-momentum tensor $(T)$, $\mathcal{L_{M}}$ represents the matter Lagrangian. Notably, we have considered the relativistic system of units, where $G=1$, and $c=1$. The matter Lagrangian is connected to the energy-momentum tensor through the relation:
\begin{equation}
	T_{ij}=-\frac{2}{\sqrt{-g}}\frac{\delta (\sqrt{-g}\mathcal{L_{M}})}{\delta g^{ij}}. \label{eq2}
\end{equation}
If the matter Lagrangian $\mathcal{L_{M}}$ depends only on the metric components $g_{ij}$, then equation~\ref{eq2} reduces to the form \cite{Landau}:
\begin{equation}
	T_{ij}=g_{ij}\mathcal{L_{M}}-2\frac{\partial \mathcal{L_{M}}}{\partial g^{ij}}. \label{eq3}
\end{equation}
By varying the action in equation~\ref{eq1} with respect to the metric tensor $g_{ij}$, one obtains the modified field equations in $f(R,T)$ gravity as:
\begin{equation}
	f_{R}(R,T)R_{ij}-\frac{1}{2}f(R,T)g_{ij}+(g_{ij}\Box-\nabla_{i}\nabla_{j})f_{R}(R,T)=8\pi T_{ij}-f_{T}(R,T)T_{ij}-f_{T}(R,T)\Theta_{ij}, \label{eq4}
\end{equation}
where $f_{R}(R,T)=\frac{\partial f(R,T)}{\partial R}$ and $f_{T}(R,T)=\frac{\partial f(R,T)}{\partial T}$. Here, $R_{ij}$ denotes the Ricci tensor, while $\Box$ represents the d'Alembert operator defined as $\Box\equiv~\frac{\partial_{i}(\sqrt{-g}g^{ij}\partial_{j})}{\sqrt{-g}}$. The operator $\nabla_{j}$ stands for the covariant derivative with respect to the Levi-Civita connection associated with $g_{ij}$, and $\Theta_{ij}=g^{\mu\nu}\frac{\delta T_{\mu\nu}}{\delta g^{ij}}$. 

The covariant divergence of equation~\ref{eq3} yields,
\begin{equation}
	\nabla^{i}T_{ij}=\frac{f_{T}(R,T)}{8\pi-f_{T}(R,T)}\Big[(T_{ij}+\Theta_{ij})\nabla^{i}~ln~f_{T}(R,T)+\nabla^{i}\Theta_{ij}-\frac{1}{2}g_{ij}\nabla^{i}T\Big]. \label{eq5}
\end{equation}  
equation~\ref{eq5} describes the non-conservation of energy-momentum tensor in the framework of $f(R,T)$ gravity. Now, the present analysis is deeply rooted in the calculation of $\Theta_{ij}$. Considering equation~\ref{eq3}, we obtain:
\begin{eqnarray}
	\frac{\delta T_{\mu\nu}}{\delta g^{ij}}=\frac{\delta g_{\mu\nu}}{\delta g^{ij}}\mathcal{L_{M}}+g_{\mu\nu}\frac{\partial \mathcal{L_{M}}}{\partial g^{ij}}-2\frac{\partial^{2}\mathcal{L_{M}}}{\partial g^{ij}\partial g^{\mu\nu}} \nonumber \\
	\;\;\;\;\;\;\;\;=\frac{\delta g_{\mu\nu}}{\delta g^{ij}}\mathcal{L_{M}}+\frac{1}{2}g_{\mu\nu}g_{ij}\mathcal{L_{M}}-\frac{1}{2}g_{\mu\nu}T_{ij}-2\frac{\partial^{2}\mathcal{L_{M}}}{\partial g^{ij}\partial g^{\mu\nu}}.\label{eq6}
\end{eqnarray}
Using the identity, $g_{\mu\eta}g^{\eta\nu}=\delta^{\nu}_{\mu}$, we reformulate the following:
\begin{equation}
	\frac{\delta g_{\mu\nu}}{\delta g^{ij}}=--g_{\mu\eta}g_{\nu\gamma}\frac{\delta g^{\eta\gamma}}{\delta g^{ij}}. \label{eq7}
\end{equation}
Now, from the definition, $\Theta_{ij}=g^{\mu\nu}\frac{\delta T_{\mu\nu}}{\delta g^{ij}}$, we obtain:
\begin{equation}
	\Theta_{ij}=-2T_{ij}+g_{ij}\mathcal{L_{M}}-2g^{\mu\nu}\frac{\partial^{2}\mathcal{L_{M}}}{\partial g^{ij}g^{\mu\nu}}. \label{eq8}
\end{equation}
Here, we adhere to the consideration that $\mathcal{L_{M}}$ depends on the metric algebraically, and linearly. Consequently, the second partial derivative on the right hand side of equation~\ref{eq8} vanishes. Hence, we get,
\begin{equation}
	\Theta_{ij}=-2T_{ij}+g_{ij}\mathcal{L_{M}}. \label{eq9}
\end{equation} 
Notably, equation~\ref{eq9} is $\mathcal{L_{M}}$ dependent. Now, we consider a well-acknowledged class of $f(R,T)$ models \cite{Harko,Pretel}, $f(R,T)=R+2\alpha_{c}T$, where, $R$, and $T$ are Ricci scalar and trace of energy-momentum tensor, and $\alpha_{c}$ is a dimensionless parameter that characterises the strength of the gravity-matter coupling. Substituting this form into equation~\ref{eq4}, we obtain the tractable form of modified field equations as,
\begin{equation}
	G_{ij}=8\pi T_{ij}+g_{ij}\alpha_{c}T-2\alpha_{c}(T_{ij}+\Theta_{ij}). \label{eq9a}
\end{equation} 
\section{Obtaining energy-momentum tensor for $\mathcal{L_{M}}=p$, and $\mathcal{L_{M}}=-\rho$}\label{sec3} 
To begin with, we consider the particle number density current in the form, $J^{i}=\sqrt{-g}nu^{i}$, and the normalisation yields, $u^{i}u^{j}g_{ij}=-1$, where, $u^{i}$ is the four-velocity. Again, $J^{i}$ can be treated as fixed, when variations with respect to metric are considered. This feature ensures that the fields of fluid remain fixed. Notably, the framework considered here, is isentropic in nature. Now, 
\begin{equation}
	g_{ij}J^{i}J^{j}=g_{ij}(\sqrt{-g})^{2}n^{2}u^{i}u^{j}=gn^{2}. \label{eq10}
\end{equation} 
Equivalently, equation~\ref{eq10} came be written as:
\begin{equation}
	n^{2}=-\frac{g_{ij}J^{i}J^{j}}{|g|}. \label{eq11}
\end{equation}
Notably, the negative sign in equation~\ref{eq11} comes from $g<0$ for Lorentzian signature, and $g_{ij}J^{i}J^{j}<0$, since $J^{i}$ is time-like, and parallel to $u^{i}$. Considering the variation variation on both sides of equation~\ref{eq11}, we obtain:
\begin{equation}
	2n\delta{n}=-\frac{1}{g}J^{i}J^{j}\delta{g_{ij}}-g_{ij}J^{i}J^{j}\delta{\Big(\frac{1}{g}\Big)}. \label{eq12}
\end{equation}
Again, 
\begin{equation}
	\delta{\Big(\frac{1}{g}\Big)}=-\frac{\delta{g}}{g^{2}}=-\frac{1}{g^{2}}gg^{\mu\nu}\delta g_{\mu\nu}. \label{eq13}
\end{equation}
Plugging equation~\ref{eq13} back into equation~\ref{eq12}, we obtain:
\begin{equation}
	2n\delta{n}=-\frac{1}{g}J^{i}J^{j}\delta{g_{ij}}+\Big(g_{ij}J^{i}J^{j}\Big)\frac{1}{g}g^{\mu\nu}\delta g_{\mu\nu}. \label{eq14}
\end{equation}
Now, using the form of number density current into equation~\ref{eq14}, and rearranging, we get,
\begin{equation}
	\delta{n}=\frac{n}{2}\Big(u^{i}u^{j}\delta{g_{ij}}+g^{\mu\nu}\delta g_{\mu\nu}\Big). \label{eq15}
\end{equation}
Using the identities, $\delta{g_{\mu\nu}}=-g_{\mu i}g_{\nu j}\delta{g^{ij}}$, and $g^{\mu\nu}\delta{g_{\mu\nu}}=-g_{\mu\nu}\delta{g^{\mu\nu}}$, we can now reformulate equation~\ref{eq15} in the form:
\begin{equation}
	\delta{n}=-\frac{n}{2}\Big(u_{i}u_{j}+g_{ij}\Big)\delta{g^{ij}}. \label{eq16}
\end{equation}
Since, the pressure and energy density depend on the particle number density $(n)$, i.e., $\rho=\rho(n)$, and $p=p(n)$, the first law of thermodynamics implies the following:
\begin{eqnarray}
	\rho=\rho(n,s) \nonumber\\
	\Rightarrow d\rho=\frac{\partial \rho}{\partial n}dn+\frac{\partial \rho}{\partial s}ds \nonumber\\
	\Rightarrow d\rho=\frac{\partial \rho}{\partial n}dn+nTds, \label{eq17}
\end{eqnarray} 
where, $T$ is the temperature. Now, for an isentropic nature $(ds=0)$, $d\rho=\frac{\partial \rho}{\partial n}dn$. Again for fixed entropy, the Gibbs-Duhem relation is expressed as:
\begin{eqnarray}
	p=n\Big(\frac{\partial\rho}{\partial n}\Big)_{s}-\rho \nonumber\\
	\Rightarrow\Big(\frac{\partial\rho}{\partial n}\Big)_{s}=\frac{\rho+p}{n}. \label{eq18}
\end{eqnarray}
Notably, for a barotropic form, and isentropic fluid, $\frac{dp}{dn}=\frac{\rho+p}{n}$ is also valid. Additionally, it must be noted that in the non-densitised current is expressed as, $N^{i}=nu^{i}$, and it is fixed under metric variations. Now, the normalisation condition reads, $n^{2}=-g_{ij}N^{i}N^{j}$. Now, repeating the steps, from equation~\ref{eq12} to equation~\ref{eq16}, we obtain:
\begin{equation}
	\delta{n}=-\frac{n}{2}u_{i}u_{j}\delta{g^{ij}}. \label{eq19}
\end{equation} 
\begin{itemize}
	\item {\bf $\mathcal{L_{M}}=p:$} Using the relation, $\frac{dp}{dn}=\frac{\rho+p}{n}$, and equation~\ref{eq19}, we obtain:
	\begin{equation}
		\frac{\partial p}{\partial g^{ij}}=-\frac{1}{2}(\rho+p)u_{i}u_{j}. \label{eq20}
	\end{equation} 
	Plugging $\mathcal{L_{M}}=p$ in the definition of energy-momentum tensor, as expressed in equation~\ref{eq3}, we obtain the following:
	\begin{equation}
		T_{ij}=pg_{ij}-2\frac{\partial p}{\partial g^{ij}}. \label{eq21}
	\end{equation}
	Substituting equation~\ref{eq20} into equation~\ref{eq21}, we get,
	\begin{equation}
		T_{ij}=(\rho+p)u_{i}u_{j}+pg_{ij}. \label{eq22}
	\end{equation}
	\item {\bf $\mathcal{L_{M}}=-\rho:$} Plugging $\mathcal{L_{M}}=-\rho$ in equation~\ref{eq3}, and using equations~\ref{eq16} and \ref{eq18}, we get,
	\begin{equation}
		T_{ij}=-\rho g_{ij}+(\rho+p)(g_{ij}+u_{i}u_{j})=(\rho+p)u_{i}u_{j}+pg_{ij}. \label{eq23}
	\end{equation} 
\end{itemize}
Following the results expressed in equations~\ref{eq20} and \ref{eq23}, we can solidify the claim that $\mathcal{L_{M}}=p$, and $\mathcal{L_{M}}=-\rho$ produce the same expression for the energy-momentum tensor as shown in equations~\ref{eq22} and \ref{eq23}, respectively.
\section{Formulating TOV equations}\label{sec4}
A static spherically symmetric space-time is characterised by the line element of the form:
\begin{equation}
	ds^{2}=-e^{2\nu}dt^{2}+e^{2\lambda}dr^{2}+r^{2}(d\theta^2+sin^{2}\theta d\phi^{2}). \label{eq24}
\end{equation}
Now, to construct the TOV equations, we begin with the modified field equations expressed in equation~\ref{eq9a}, and make use of the standard result, $e^{-2\lambda}=1-\frac{2m(r)}{r}$. From equation~\ref{eq9}, we observe that $\Theta_{ij}$ is $\mathcal{L_{M}}$ dependent, and this dependence influences the formulation of Tolman-Oppenheimer-Volkoff equation \cite{Tolman,Oppenheimer}. Using the above framework, we now proceed toward calculating the TOV equations \cite{Tolman, Oppenheimer} for both the choices $\mathcal{L_{M}}=p$, and $\mathcal{L_{M}}=-\rho$.  
\begin{itemize}
	\item {\bf Case-I:} Substituting $\mathcal{L_{M}}=p$ in equation~\ref{eq9}, we obtain, $\Theta_{ij}=-2T_{ij}+pg_{ij}$, or $\Theta=-2T+4p$. This form is widely accepted and used in almost every research concerning $f(R,T)$ framework \cite{Harko, Pretel}. In this context, the conservation of energy-momentum tensor, expressed in equation~\ref{eq5}, yields, $\nabla^{i}T_{ij}=\frac{\alpha_{c}}{8\pi+2\alpha_{c}}(\rho'-p')$ \cite{Pretel}. Consequently, the modified TOV equations take the following form:
	\begin{equation}
		\frac{dm}{dr}=4\pi r^{2}\rho+\frac{\alpha_{c} r^{2}}{2}\Big[3\rho-p\Big], \label{eq25}
	\end{equation} 
	and
	\begin{equation}
		\frac{dp}{dr}=\Bigg[\frac{\alpha_{c}}{8\pi+3\alpha_{c}}\Bigg]\rho'-\Bigg[\frac{8\pi+2\alpha_{c}}{8\pi+3\alpha_{c}}\Bigg](\rho+p)\Bigg[\frac{m(r)+4\pi r^{3}p_{eff}}{r[r-2m(r)]}\Bigg], \label{eq26}
	\end{equation}
	where, $p_{eff}=p+\frac{\alpha_{c}}{8\pi}(3p-\rho)$.
	\item {\bf Case-II:} Similarly, using $\mathcal{L_{M}}=-\rho$ in equation~\ref{eq9} yields, $\Theta_{ij}=-2T_{ij}-\rho g_{ij}$, or $\Theta=-2T-4\rho$. Further, in this context, covariant divergence of energy-momentum generates, $\nabla^{i}T_{ij}=\frac{\alpha_{c}}{8\pi+2\alpha_{c}}(-\rho'-3p')$. Hence, the TOV equation takes the following form:
	\begin{equation}
		\frac{dm}{dr}=\frac{\rho r^{2}}{2}(8\pi+\alpha_{c})-\frac{3}{2}\alpha_{c}pr^{2}, \label{eq27}
	\end{equation}
	and
	\begin{equation}
		\frac{dp}{dr}=-\Bigg[\frac{\alpha_{c}}{8\pi+5\alpha_{c}}\Bigg]\rho'-\Bigg[\frac{8\pi+2\alpha_{c}}{8\pi+5\alpha_{c}}\Bigg](\rho+p)\Bigg[\frac{m(r)+4\pi r^{3}p_{eff}}{r[r-2m(r)]}\Bigg], \label{eq28}
	\end{equation}
	where, $p_{eff}=p+\frac{\alpha_{c}}{8\pi}(5p+\rho)$.
\end{itemize}
Clearly, the set of equations~\ref{eq25} and \ref{eq26} for the choice $\mathcal{L_{M}}=p$ are different from the set of equations~\ref{eq27} and \ref{eq28} for the choice $\mathcal{L_{M}}=-\rho$.
\section{M-R diagrams}\label{sec5}
In this section, we determine the maximum mass and corresponding radius achievable by a compact star. To do so, we need to solve the TOV equations using some EoS. In practice, there are a large number of EoS available, which describe different aspects of stellar configurations. In the present analysis, we consider the MIT bag model EoS \cite{Chodos,Farhi}, {\it viz.}, $p=\frac{1}{3}(\rho-4B_{g})$, where, $B_{g}$ is the bag constant, signifying the difference between vacuum energies pertaining to the perturbative and non-perturbative nature. Following the work of Madsen \cite{Madsen}, we restrict the choice of $B_{g}$ within the range $57.55\leq B_{g}\leq95.11~MeV/fm^{3}$. Notably, in the limit of massless strange quarks, this range of $B_{g}$ supports stable strange matter \cite{Madsen}. This EoS has been greatly influential in describing the interior properties of compact stars. Additionally, for a comprehensive analysis, we have varied $\alpha_{c}$ in the range from $-2$ to $2$, following previous studies \cite{Deb,Carvalho}. Considering this setup, we solve the TOV equations, as outlined in the previous section. The resulting mass-radius (M-R) relations are analysed in the following cases:
\begin{itemize}
	\item {\bf Case-I:} We solve TOV equations~\ref{eq25} and \ref{eq26} pertaining to $\mathcal{L_{M}}=p$, and obtain the M-R plot shown in \ref{fig1}. From \ref{fig1}, the maximum masses are evaluated and are tabulated in tab.~\ref{tab1} for different values of $\alpha_{c}$. 
	\begin{figure}[h]
		\begin{subfigure}{0.5\textwidth}
			\centering
			\includegraphics[width=\textwidth]{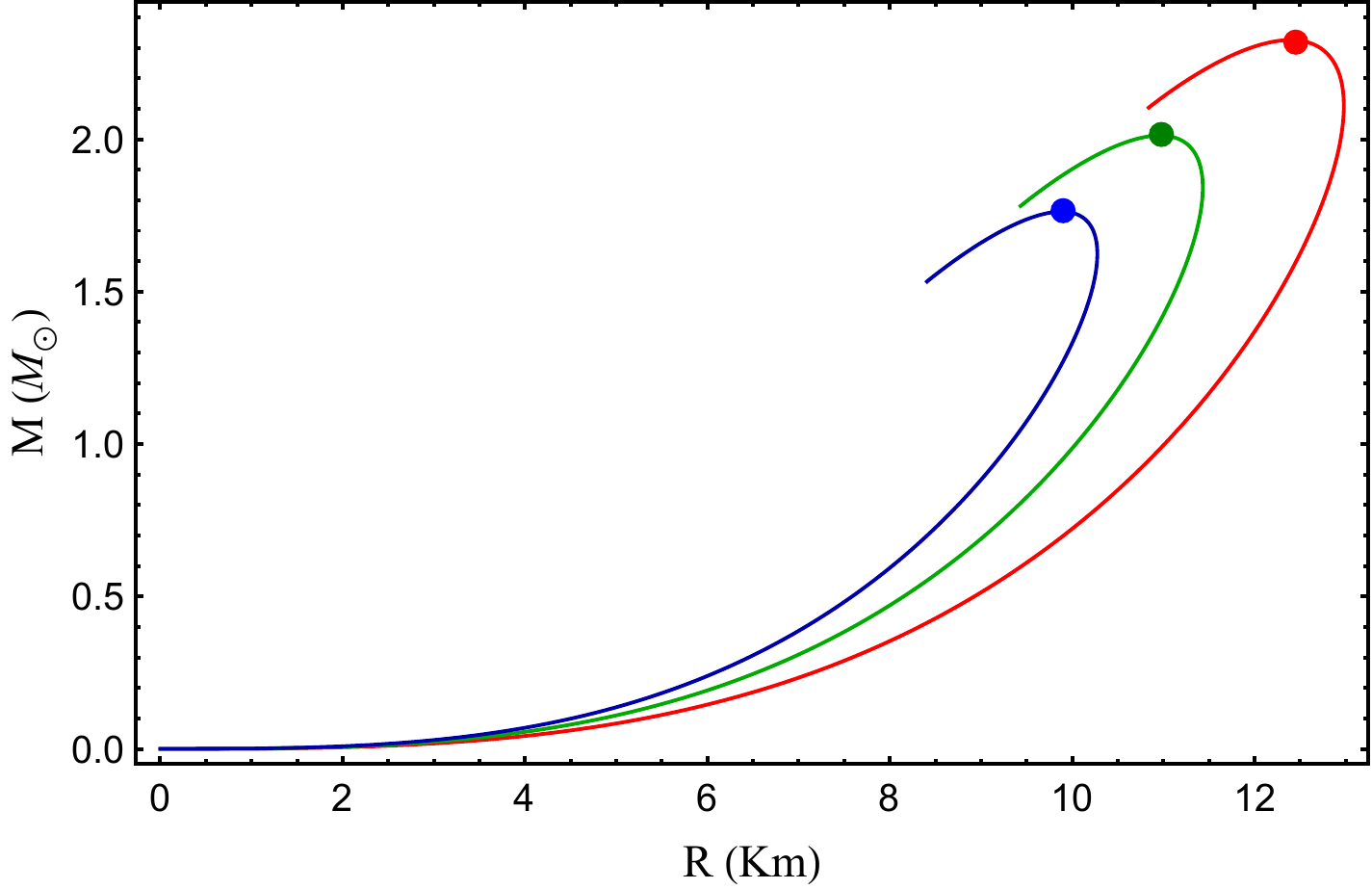}
			\caption{(a)}
			\label{fig1a}
		\end{subfigure}
		\hfil
		\begin{subfigure}{0.5\textwidth}
			\centering
			\includegraphics[width=\textwidth]{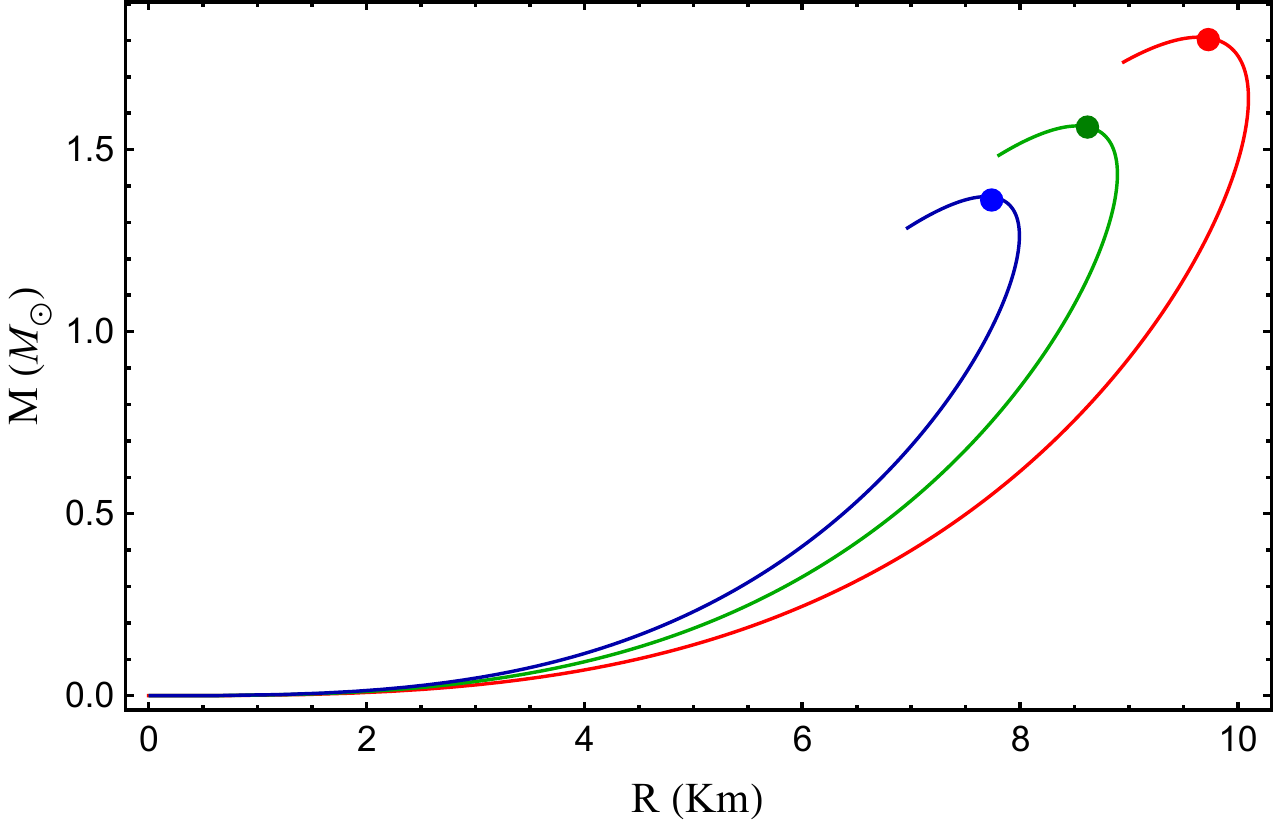}
			\caption{(b)}
			\label{fig1b}
		\end{subfigure}
		\caption{M-R plots for (a) $B_{g}=57.55~MeV/fm^{3}$, and (b) $B_{g}=95.11~MeV/fm^{3}$. Here, red, green, and blue lines are for $\alpha_{c}=-2,~0$, and $2$, respectively, and the solid dots represent the associated maximum mass points.}
		\label{fig1}
	\end{figure}
	\begin{table}[h]
		\centering
		\begin{tabular}{cc}
			\begin{minipage}{0.5\linewidth}
				\centering
				\begin{tabular}{ccc}
					\hline
					$\alpha_{c}$ & Maximum mass & Radius\\
					& $\mathrm{(M_{\odot})}$ & (Km) \\
					\hline
					-2 & 2.32 & 12.39 \\
					0 & 2.012 & 10.96 \\
					2 & 1.76 & 9.89 \\
					\hline
				\end{tabular}
				\subcaption{(a)}
				\label{tab1a}
			\end{minipage}
			\hfil
			\begin{minipage}{0.5\linewidth}
				\centering
				\begin{tabular}{ccc}
					\hline
					$\alpha_{c}$ & Maximum mass & Radius\\
					& $\mathrm{(M_{\odot})}$ & (Km) \\
					\hline
					-2 & 1.81 & 9.64 \\
					0 & 1.56 & 8.53 \\
					2 & 1.37 & 7.69 \\
					\hline
				\end{tabular}
				\subcaption{(b)}
				\label{tab1b}
			\end{minipage}
		\end{tabular}
		\caption{Tabulation of maximum mass and radius, as obtained from \ref{fig1}, for (a) $B_{g}=57.55~MeV/fm^{3}$, and (b) $B_{g}=95.11~MeV/fm^{3}$.}
		\label{tab1}
	\end{table}
	From \ref{fig1}, we observe that for a particular $B_{g}$, with increasing $\alpha_{c}$, the maximum mass decreases. This implies that with increasing gravity-matter coupling, the effective strength of gravity increases. As a result, counter-balancing the gravitational collapse becomes harder for the pressure component. Consequently, the system tends to supports less massive structure. Moreover, in a comparative scenario between tables~\ref{tab1a} and \ref{tab1b}, we note that for a particular $\alpha_{c}$, increasing $B_{g}$ results in a decrease in the maximum mass-radius. This feature stems from the EoS in the following way: with increasing $B_{g}$, the pressure decreases. As a result, the EoS becomes softer, and the compressibility of the internal matter distribution increases, leading to a smaller maximum mass. Moreover, we have constrained the parameter space through the solution of TOV equation. We observe that for $B_{g}=57.55~MeV/fm^{3}$, the allowed range of the coupling parameter $(\alpha_{c})$ lies between a lower bound of $\alpha_{c}=-4.3$ and an upper bound defined by $\alpha_{c}=15.5$. The maximum mass corresponding to $\alpha_{c}=-4.3$ is $2.78~M_{\odot}$, and corresponding radius is $14.86~Km$. Similarly, for $B_{g}=95.11~MeV/fm^{3}$, the parameter $\alpha_{c}$ is bounded between $-5.5$ and $9.4$. The associated maximum mass is $2.39~M_{\odot}$ for $\alpha_{c}=-5.5$, and corresponding radius is $12.99~Km$. Beyond these specified bounds of $\alpha_{c}$, the solution of TOV equations does not provide physically viable results.
	\item {\bf Case-II:} Here, we consider equations~\ref{eq27} and \ref{eq28} associated with $\mathcal{L_{M}}=-\rho$. The solution of TOV equation in this context yields the the M-R plot shown in \ref{fig2}. Maximum masses are calculated from \ref{fig2} and are tabulated in tab.~\ref{tab2} for different $\alpha_{c}$. 
	\begin{figure}[h]
		\begin{subfigure}{0.5\textwidth}
			\centering
			\includegraphics[width=\textwidth]{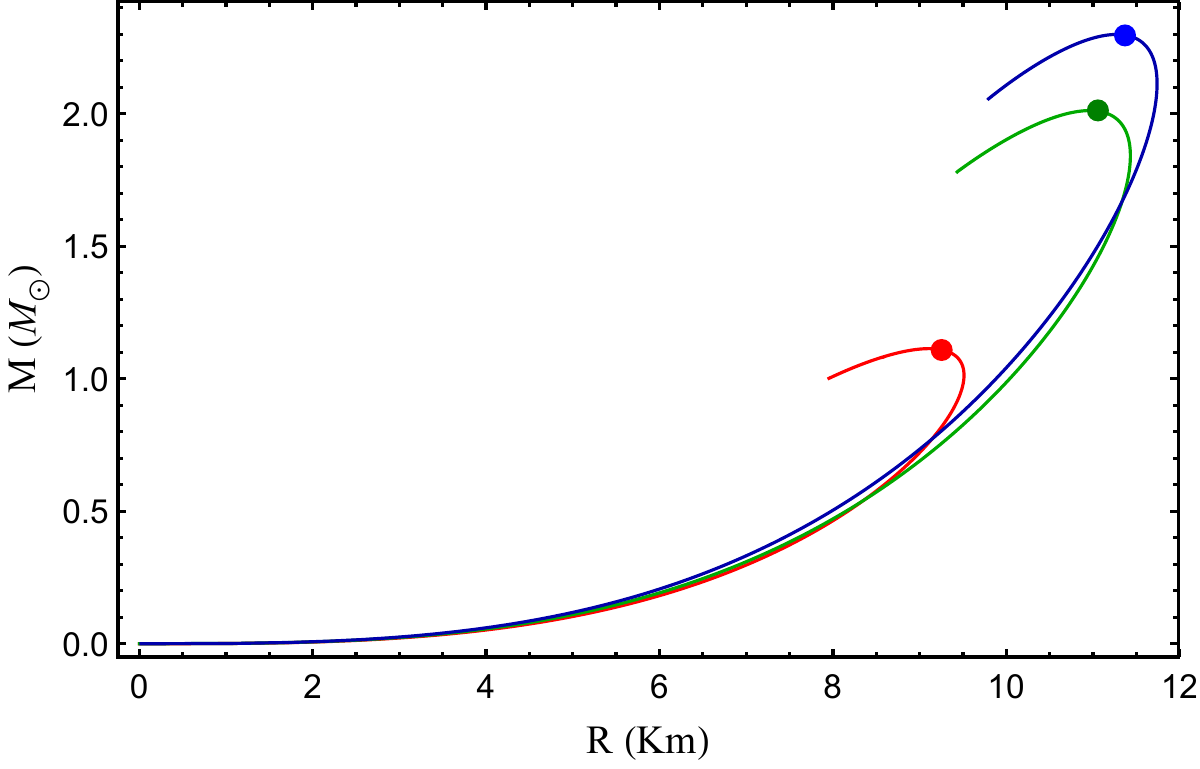}
			\caption{(a)}
			\label{fig2a}
		\end{subfigure}
		\hfil
		\begin{subfigure}{0.5\textwidth}
			\centering
			\includegraphics[width=\textwidth]{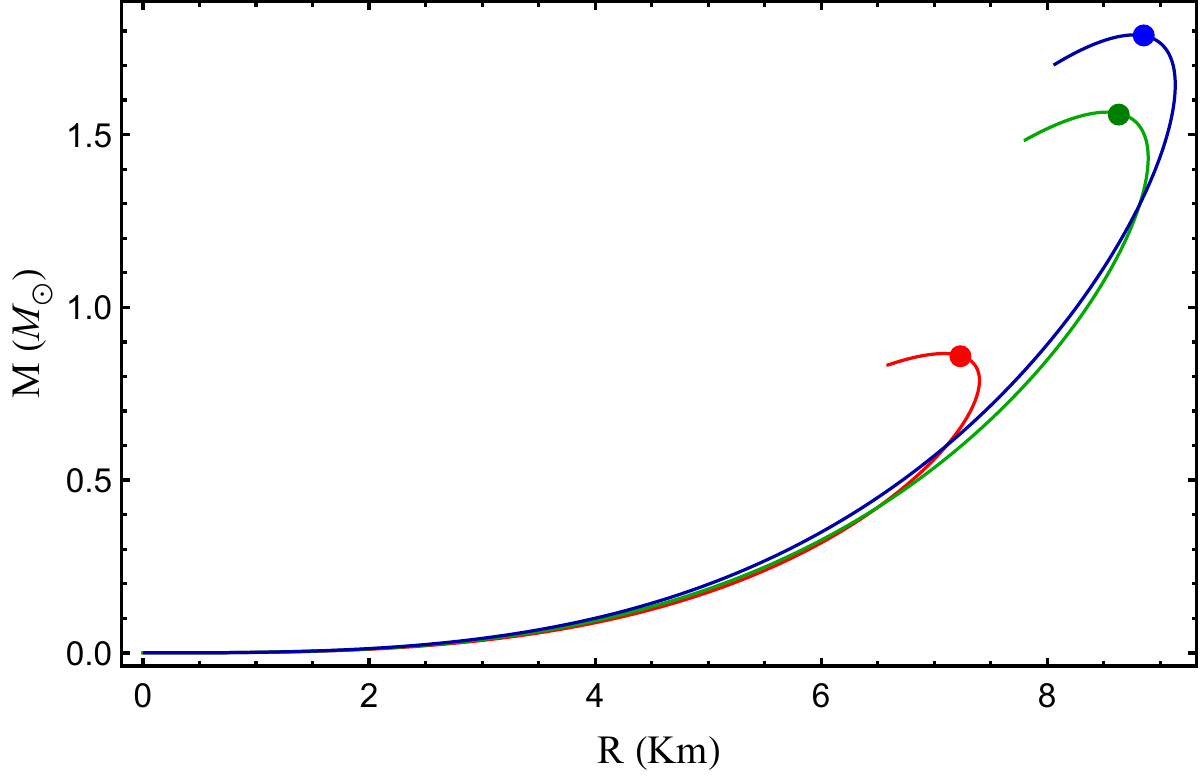}
			\caption{(b)}
			\label{fig2b}
		\end{subfigure}
		\caption{M-R plots for (a) $B_{g}=57.55~MeV/fm^{3}$, and (b) $B_{g}=95.11~MeV/fm^{3}$. Here, red, green, and blue lines are for $\alpha_{c}=-2,~0$, and $2$, respectively, and the solid dots represent the associated maximum mass points.}
		\label{fig2}
	\end{figure}
	\begin{table}[h]
		\centering
		\begin{tabular}{cc}
			\begin{minipage}{0.5\linewidth}
				\centering
				\begin{tabular}{ccc}
					\hline
					$\alpha_{c}$ & Maximum mass & Radius\\
					& $\mathrm{(M_{\odot})}$ & (Km) \\
					\hline
					-2 & 1.11 & 9.11 \\
					0 & 2.012 & 10.96 \\
					2 & 2.30 & 11.28 \\
					\hline
				\end{tabular}
				\subcaption{(a)}
				\label{tab2a}
			\end{minipage}
			\hfil
			\begin{minipage}{0.5\linewidth}
				\centering
				\begin{tabular}{ccc}
					\hline
					$\alpha_{c}$ & Maximum mass & Radius\\
					& $\mathrm{(M_{\odot})}$ & (Km) \\
					\hline
					-2 & 0.86 & 7.08 \\
					0 & 1.56 & 8.53 \\
					2 & 1.79 & 8.77 \\
					\hline
				\end{tabular}
				\subcaption{(b)}
				\label{tab2b}
			\end{minipage}
		\end{tabular}
		\caption{Tabulation of maximum mass and radius, as obtained from \ref{fig2}, for (a) $B_{g}=57.55~MeV/fm^{3}$, and (b) $B_{g}=95.11~MeV/fm^{3}$.}
		\label{tab2}
	\end{table}
	In comparison to case-I, we observe a different trend in the M-R plots here. In this scenario, with a fixed $B_{g}$ and increasing $\alpha_{c}$, the maximum mass and radius increase. This behaviour may be related to the fact that the choice, $\mathcal{L_{M}}=-\rho$ introduces modifications in the structure of TOV equations, and this modified dynamics induces effective pressure gradients. The combined effect of all these factors may cause the system to support more mass with increasing gravity-matter coupling. However, for a particular $\alpha_{c}$, the maximum mass decreases with increasing $B_{g}$, which is an well-established result, as discussed earlier. Notably, we have obtained the bounds on $\alpha_{c}$ in this context too. For $B_{g}=57.55~MeV/fm^{3}$, the admissible range for the coupling parameter $(\alpha_{c})$ extends from a lower bound of $\alpha_{c}=-2.87$  to an upper bound, as high as $\alpha_{c}=56.6$. In particular, for $B_{g}=57.55~MeV/fm^{3}$, the maximum mass achieved in this context is $2.41~M_{\odot}$ with a radius of $11.07~Km$ at $\alpha_{c}=6$. Similarly, for $B_{g}=95.11~MeV/fm^{3}$, the parameter $\alpha_{c}$ is  constrained within the range $-2.7\leq\alpha_{c}\leq29$. In this regime, the maximum mass obtained is $1.87~M_{\odot}$ with a radius of $8.61~Km$, for $\alpha_{c}=6$. Notably, beyond the specified bounds of $\alpha_{c}$ in either case, the solution to the TOV equations ceases to yield physically viable stellar configurations. In the case of $\mathcal{L_{M}}=-\rho$, we note one interesting result. When $\alpha_{c}$ increases from allowed minimum value, the maximum mass initially increases, attains a maximum value, and then decreases. To illustrate this behaviour, the dependence of maximum mass on $\alpha_{c}$ is presented in Figures~\ref{fig3a} and \ref{fig3b}. This nature may be explained in the following way: in the present context, when $\alpha_{c}$ is finite but small, the effects of modified gravity are mild. Now, increasing $\alpha_{c}$ may provide additional support against gravitational collapse, so the star can accommodate more mass. Hence, the maximum mass increases. As $\alpha_{c}$ continues to grow, the gravitational modifications may become stronger, and often non-linear. They may begin to enhance gravity rather than suppress it, leading to instabilities. Consequently, the internal pressure can no longer effectively counteract gravity, resulting in a decrease in the maximum mass.   
	\begin{figure}[h]
		\begin{subfigure}{0.5\textwidth}
			\centering
			\includegraphics[width=\textwidth]{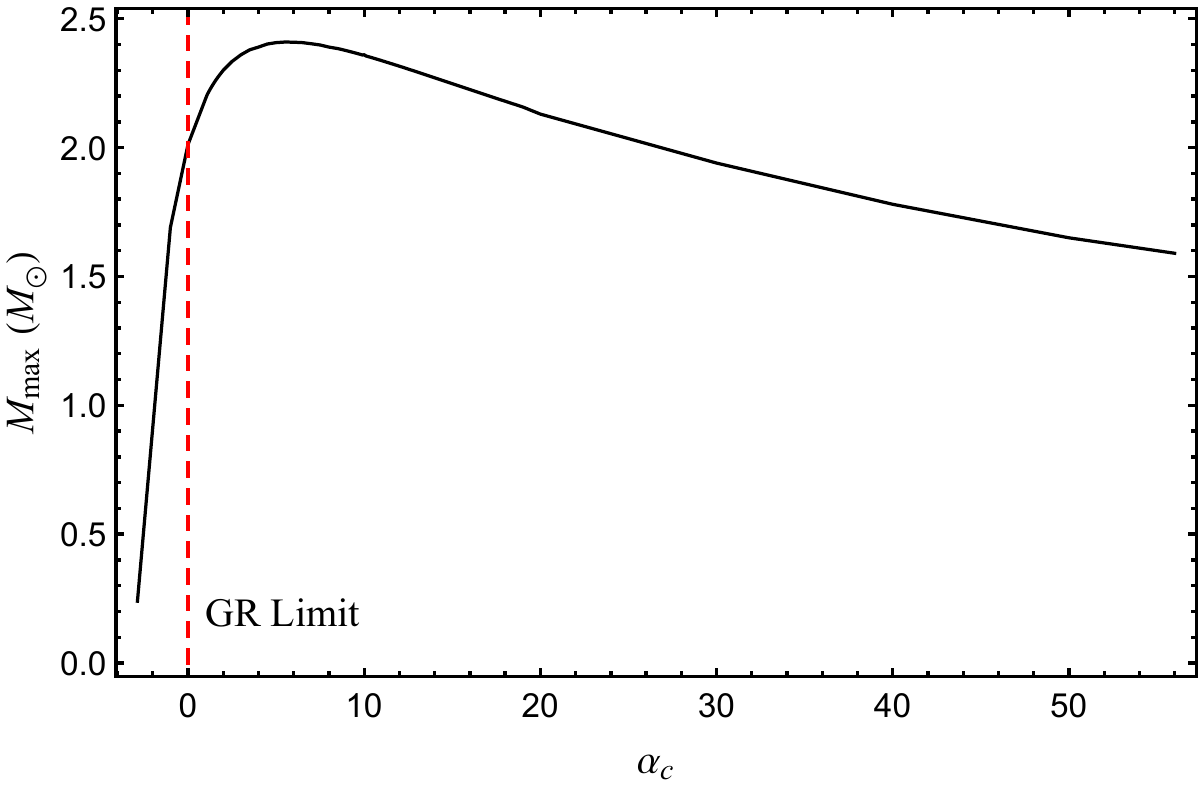}
			\caption{(a)}
			\label{fig3a}
		\end{subfigure}
		\hfil
		\begin{subfigure}{0.5\textwidth}
			\centering
			\includegraphics[width=\textwidth]{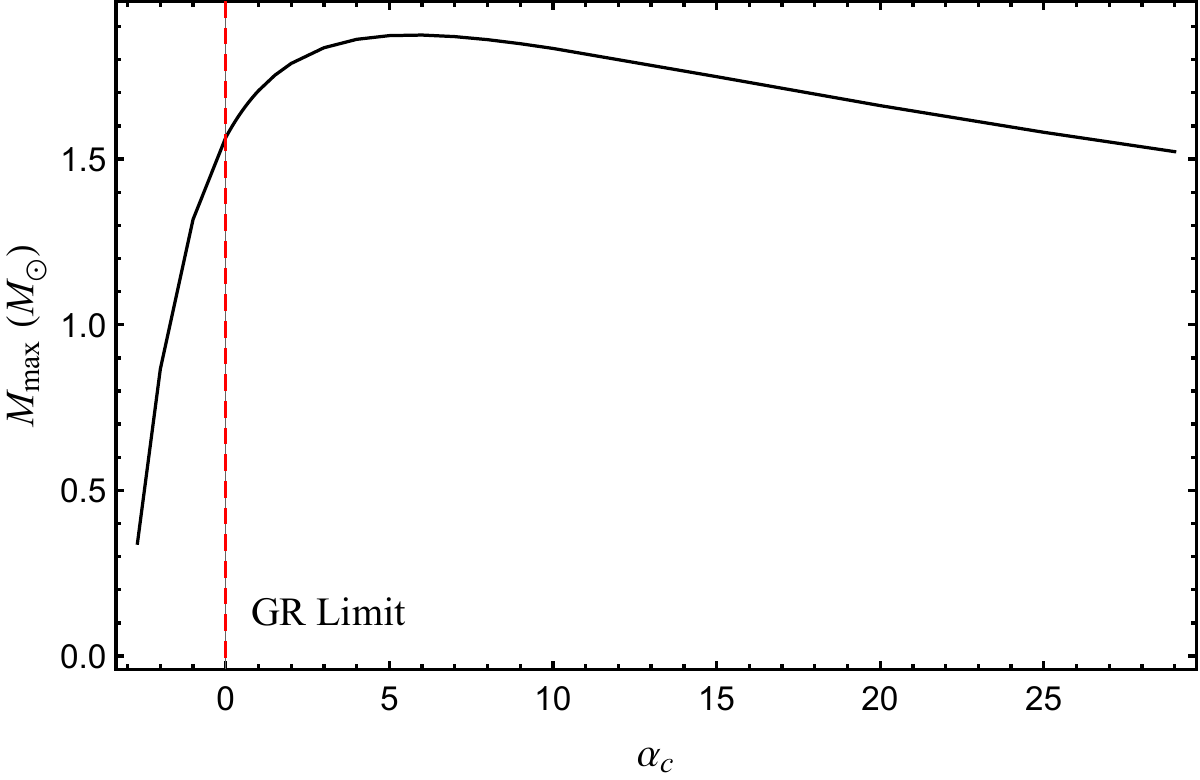}
			\caption{(b)}
			\label{fig3b}
		\end{subfigure}
		\caption{$M_{max}$ vs $\alpha_{c}$ plots for (a) $B_{g}=57.55~MeV/fm^{3}$, and (b) $B_{g}=95.11~MeV/fm^{3}$.}
		\label{fig3}
	\end{figure}			
\end{itemize}
\section{Discussion}\label{sec6}
The proper choice of matter Lagrangian in $f(R,T)$ theory of gravity, has been a prolonged and unresolved issue. In the pioneering work of Harko et al. \cite{Harko}, it was mentioned that $\mathcal{L_{M}}=p$, and $\mathcal{L_{M}}=-\rho$ both produce the same expression of the energy-momentum tensor. In spite of this, the ambiguity in the definition of the energy-momentum tensor persists. In the present paper, we have presented the in-depth calculation, from thermodynamical point of view, to show that both the choices of $\mathcal{L_{M}}$ produce the same energy-momentum tensor, indeed. Starting from the very point of densitised and non-densitised current densities, the GR and thermodynamic treatment produced equations~\ref{eq22} and \ref{eq23} proving that the matter Lagrangian $\mathcal{L_{M}}=p$, and $\mathcal{L_{M}}=-\rho$ yield the exact same result. Hence, the conjecture is proven. Now, to assess the physical viability of these choices, we start by deriving the modified field equations, within the framework of $f(R,T)$ gravity, in equation~\ref{eq4}. Refinement of equation~\ref{eq4} through different mathematical intricacies leads to the summarised field equations of equation~\ref{eq9a}. It must be noted that the modified field equations are highly dependent on the choice of $\mathcal{L_{M}}$, through $T_{ij}$, and $\Theta_{ij}$. Now, using equation~\ref{eq9a}, and a static spherically symmetric space-time, we have derived the TOV equations for both $\mathcal{L_{M}}=p$, and $\mathcal{L_{M}}=-\rho$, in equations~\ref{eq25}-\ref{eq26}, and \ref{eq27}-\ref{eq28}, respectively. Next, we solve the TOV equations using MIT bag model EoS. Notably, following the work of Madsen \cite{Madsen}, we have varied the bag parameter $(B_{g})$ in the range $57.55\leq B_{g}\leq95.11~MeV/fm^{3}$, in accordance with stable strange matter hypothesis. For both the choices of $\mathcal{L_{M}}$, the results are illustrated in figures~\ref{fig1}, and \ref{fig2}, and the concerned maximum mass-radius are tabulated in tables~\ref{tab1}, and \ref{tab2}. It must be observed that the M-R plots are obtained for the lower and upper bound of $B_{g}$. Interestingly, we have noted that the nature of the M-R plots for $\mathcal{L_{M}}=p$ are different to that for $\mathcal{L_{M}}=-\rho$. For instance, the solution of TOV in the framework of $f(R,T)$ considering $\mathcal{L_{M}}=p$, shows that when $B_{g}$ is fixed, increasing gravity-matter coupling $(\alpha_{c})$ decreases the maximum mass \cite{Pretel,Carvalho}. This can be attributed to the fact that with increasing $\alpha_{c}$, the strength of effective gravity increases. As a result the compact star's tendency of gravitational collapse increases. Further, the modifications introduced by this choice act as effective gravitational softening of the EoS, yielding a lower value of maximum mass. Hence, $\mathcal{L_{M}}=p$ may be termed attractive correction in this framework. Now, for $\mathcal{L_{M}}=-\rho$, the maximum mass increases with $\alpha_{c}$ for a particular $B_{g}$. The choice of $\mathcal{L_{M}}=-\rho$ induce effective pressure gradients and forces, which may be the reason to allow for a greater maximum mass. In this way $\mathcal{L_{M}}=-\rho$ may be termed repulsive correction in this framework. Further, from MIT bag model EoS, it is notable that with increasing $B_{g}$, the EoS becomes softer, and consequently, the maximum mass decreases. This feature is however preserved in both the choices of $\mathcal{L_{M}}$. We have also obtained the bounds of $\alpha_{c}$ though the physical acceptability of the TOV solutions. Now, we aim to scrutinise the credibility of these choices under consideration. To validate them, we have predicted the radii of some recently observed pulsars, and lighter companions of gravitational wave (GW) events, for the two choices of $\mathcal{L_{M}}$, and presented a comparative study in \ref{tab3}. Notably, we have considered $B_{g}=57.55~MeV/fm^{3}$ for the radius prediction. 
\begin{table*}
	\centering
	\begin{tabular}{ccccc|cc}
		\hline
		\multirow{3}{*}{Compact objects} & \multirow{3}{*}{Observed mass} & \multirow{3}{*}{Measured radius} & \multicolumn{4}{c}{Predicted radius from TOV} \\ \cline{4-7}
		&  &  & \multicolumn{2}{c}{$\mathcal{L_{M}}=p$}&\multicolumn{2}{c}{$\mathcal{L_{M}}=-\rho$}\\ \cline{4-7}
		&$(M_{\odot})$ & $(Km)$ & $\alpha_{c}$ & R (Km)&$\alpha_{c}$ & R (Km)\\ \hline 
		\vspace{0.2cm}
		GW 190814 \cite{Abbott3} & $2.59^{+0.08}_{-0.09}$ & -- & -3.40 & 13.85 & -- & -- \\ 
		\vspace{0.2cm}
		PSR J0952-0607 \cite{Carvalho1} & $2.35^{+0.17}_{-0.17}$ & --  & -2.20 & 12.79 & 3.00 & 11.40 \\
		\vspace{0.2cm}
		PSR J2215+5135 \cite{Linares} & $2.27^{+0.17}_{-0.15}$& -- &-1.70 & 12.28 & 1.80 & 11.48 \\
		\vspace{0.2cm}
		PSR J0740+6620 \cite{Riley} & $2.072^{+0.067}_{-0.066}$ & $12.39^{+1.30}_{-0.98}$ & -1.30 & 12.36 & 0.80 & 11.59 \\
		\vspace{0.2cm}
		PSR J1614-2230 \cite{Demorest} & $1.97^{+0.04}_{-0.04}$ & 11-15 & -0.60 & 11.83 & 0.10 & 11.39 \\
		\vspace{0.2cm}
		4U 1608-52 \cite{Guver} & $1.74^{+0.14}_{-0.14}$ & $9.3^{+1.0}_{-1.0}$ & 2.20 & 9.81 & 26.8 & 9.29\\
		\vspace{0.2cm}
		4U 1820-30 \cite{Guver1} & $1.58^{+0.06}_{-0.06}$ & $9.1^{+0.4}_{-0.4}$ & 3.78 & 9.15 & 26.5 & 9.13\\
		\hline
	\end{tabular}
	\caption{Radius prediction from both $\mathcal{L_{M}}=p$ and $\mathcal{L_{M}}=-\rho$}
	\label{tab3}
\end{table*}  
From \ref{tab3}, it is clear that adopting $\mathcal{L_{M}}=p$ provides a more suitable framework for predicting the radii of different compact stars as well as lighter mass of GW events, GW 190814 \cite{Abbott3}. On the other hand, choosing $\mathcal{L_{M}}=-\rho$ fails to produce a maximum mass of $2.59~M_{\odot}$, thereby restricting its applicability to regime of high-mass compact stars. With the rapid progress in astronomical observations, it is likely that compact stars significantly more massive than $2.59~M_{\odot}$ will be detected, in which case the $\mathcal{L_{M}}=-\rho$ choice would lead to theoretical modeling with less applicability. Consequently, since $\mathcal{L_{M}}=p$ demonstrates greater versatility in capturing the mass-radius relations across a wider class of compact stars, it may serve as a useful criterion for identifying the appropriate matter Lagrangian.

In conclusion, we emphasise that both choices of the matter Lagrangian $(\mathcal{L_{M}})$ are equally valid, but not equivalent as certain restrictions may arise when dealing with the solutions of TOV equations. While the choices $\mathcal{L_{M}}=p$, and $\mathcal{L_{M}}=-\rho$ give the same energy-momentum tensor , the form of TOV equations are different. As a result internal structures are also significantly modified for the choices of $\mathcal{L_{M}}$. Notably, the allowed range of $\alpha_{c}$ is modified for the choice of $\mathcal{L_{M}}$. This can be explained as follows: in $f(R,T)$ gravity, the range of the coupling constant differs for the choices $\mathcal{L_{M}}=p$ and $\mathcal{L_{M}}=-\rho$ because the matter Lagrangian directly influences the form of the matter-geometry coupling in the field equations. Although both Lagrangian forms yield the same energy–momentum tensor, they lead to different expressions for $\Theta_{ij}$, and hence to different coupling terms involving $f_{T}(T_{ij}+\Theta_{ij})$. For instance, in the present linear case $f(R,T)=R+2\alpha_{c}T$, the effective equations take different forms depending on the choice of $\mathcal{L_{M}}$. This difference changes the sign and strength of the geometric corrections to the effective pressure and density, thereby modifying the stability, and hydrostatic balance of stellar configurations. As a result, the physically admissible range of the coupling constant $\alpha_{c}$ differs for the two choices of $\mathcal{L_{M}}=p$ and $\mathcal{L_{M}}=-\rho$, as evident in the present study. This ultimately accounts for the maximum mass achievable as well as other gross properties of a star. Moreover, in this model we have obtained a maximum mass of $2.78~M_{\odot}$ and radius of $14.86~Km$ for $\mathcal{L_{M}}=p$, $B_{g}=57.55~MeV/fm^{3}$, $\alpha_{c}=-4.7$. On the other hand, for $\mathcal{L_{M}}=-\rho$, the maximum mass is $2.41~M_{\odot}$ with a radius of $11.07~Km$ for $B_{g}=57.55~MeV/fm^{3}$, and $\alpha_{c}=6$. Hence, the selection of a specific form of $\mathcal{L_{M}}$ ultimately depends on the researcher's approach and physical considerations. The present framework will inspire further investigations and open new avenues for exploring various aspects of compact stellar structures within the $f(R,T)$ theory of gravity.
\section*{Acknowledgements}
DB is thankful to the Department of Science and Technology (DST), Govt. of India, for providing the fellowship vide no:  DST/INSPIRE Fellowship/2021/IF210761. PKC gratefully acknowledges support from IUCAA, Pune, India under Visiting Associateship programme. DB is also grateful to IUCAA, Pune, India for providing the visitor facilities. DB and PKC gratefully acknowledge the facilities provided by IUCAA during their visit, where this work was completed.     
\section*{Data availability statement}
All data that support the findings of this study are included within the article (and any supplementary files).

\end{document}